\documentclass[sigconf]{acmart}
\AtBeginDocument{%
  }
\copyrightyear{2025}
\acmYear{2025}
\setcopyright{cc}
\setcctype{by-nc-sa}
\acmConference[SIGIR-AP 2025]{Proceedings of the 2025 Annual International ACM SIGIR Conference on Research and Development in Information Retrieval in the Asia Pacific Region}{December 7--10, 2025}{Xi'an, China}
\acmBooktitle{Proceedings of the 2025 Annual International ACM SIGIR Conference on Research and Development in Information Retrieval in the Asia Pacific Region (SIGIR-AP 2025), December 7--10, 2025, Xi'an, China}
\acmDOI{10.1145/3767695.3769484}
\acmISBN{979-8-4007-2218-9/2025/12}
\usepackage{algorithm}
\usepackage{algorithmic}
\usepackage{tikz}
\usepackage{amsmath}
\usepackage{subcaption}
\usepackage{tabularx}
\usepackage{makecell}
\usepackage{booktabs}
\usepackage{array}
\usepackage{fontawesome5}
\usepackage[most]{tcolorbox}
\definecolor{pastelgreen_hot}{RGB}{204, 100, 197}
\definecolor{pastelgreen}{RGB}{204, 235, 197}
\definecolor{brightgreen}{RGB}{0, 50, 0}
\usepackage{xcolor}
\usepackage{fontawesome5}
\definecolor{p1bubble}{RGB}{235,222,235}
\definecolor{agentbubble}{RGB}{251,229,186} 
\definecolor{replybubble}{RGB}{235,245,255} 
\definecolor{p2bubble}{RGB}{220,235,250} 
\newtcolorbox{p1chat}{
  colback=p1bubble,
  colframe=white,
  arc=6pt,
  left=2pt,
  right=2pt,
  top=2pt,
  bottom=2pt,
  boxrule=0pt,
  width=\columnwidth,
  sharp corners=southwest,
  before skip=2pt,
  after skip=2pt
}
\newtcolorbox{p2chat}{
  colback=p2bubble,
  colframe=white,
  arc=6pt,
  left=2pt,
  right=2pt,
  top=2pt,
  bottom=2pt,
  boxrule=0pt,
  width=\columnwidth,
  sharp corners=southwest,
  before skip=2pt,
  after skip=2pt
}
\newtcolorbox{agentchat}{
  colback=agentbubble,
  colframe=white,
  arc=4pt,
  left=2pt,
  right=2pt,
  top=2pt,
  bottom=2pt,
  boxrule=0pt,
  width=\columnwidth,
  sharp corners=southeast,
  before skip=2pt,
  after skip=2pt
}
\newtcolorbox{replychat}{
  colback=replybubble,
  colframe=white,
  arc=4pt,
  left=2pt,
  right=2pt,
  top=2pt,
  bottom=2pt,
  boxrule=0pt,
  width=\columnwidth,
  before skip=2pt,
  after skip=2pt
}

\begin{document}

\title{Beyond Static Evaluation: Rethinking the Assessment of Personalized Agent Adaptability in Information Retrieval}

\author{Kirandeep Kaur}
\affiliation{%
  \institution{University of Washington}
  \city{Seattle}
  \state{WA}
  \country{USA}
}
\email{kaur13@cs.washington.edu}

\author{Preetam Prabhu Srikar Dammu}
\affiliation{%
  \institution{University of Washington}
  \city{Seattle}
  \state{WA}
  \country{USA}
}
\email{preetams@uw.edu}

\author{Hideo Joho}
\affiliation{%
  \institution{University of Tsukuba}
  \city{Tsukuba}
  \country{Japan}
}
\email{joho.hideo.gb@u.tsukuba.ac.jp}

\author{Chirag Shah}
\affiliation{%
  \institution{University of Washington}
  \city{Seattle}
  \state{WA}
  \country{USA}
}
\email{chirags@uw.edu}

\renewcommand{\shortauthors}{Kirandeep Kaur, Preetam Prabhu Srikar Dammu, Hideo Joho, Chirag Shah}
\renewcommand{\shorttitle}{Beyond Static Evaluation}
\begin{abstract}
Personalized AI agents are becoming central to modern information retrieval, yet most evaluation methodologies remain static — relying on fixed benchmarks and one-off metrics that fail to reflect how users’ needs evolve over time. These limitations hinder our ability to assess whether agents can meaningfully adapt to individuals across dynamic, longitudinal interactions. In this \emph{perspective} paper, we propose a conceptual lens for rethinking evaluation in adaptive personalization — shifting the focus from static performance snapshots to interaction-aware, evolving assessments. We organize this lens around three core components: (1) persona-based user simulation with temporally evolving preference models; (2) structured elicitation protocols inspired by reference interviews to extract preferences in context; and (3) adaptation-aware evaluation mechanisms that measure how agent behavior improves across sessions and tasks. While recent works have embraced LLM-driven user simulation, we situate this practice within a broader paradigm for evaluating agents over time. To illustrate our ideas, we conduct a case study in e-commerce search using the PersonalWAB dataset. Beyond presenting a framework, our work lays a conceptual foundation for understanding and evaluating personalization as a continuous, user-centric endeavor.
\end{abstract}


\maketitle
\section{Introduction}
Information Retrieval (IR) systems have traditionally relied on static benchmarks and fixed evaluation methodologies to assess performance~\cite{kelly2009methods}. Rooted in the Cranfield paradigm, these approaches emphasized document relevance judged by the same individuals who expressed the information need~\cite{cleverdon1960cranfield}. Over time, practical constraints led to the decoupling of need expression and relevance judgment~\cite{scholer2016testcollections}, and recent developments such as LLM-as-a-judge~\cite{li2024llmjudge} have further shifted evaluation away from human users altogether.

While these adaptations offer scalability, they move evaluation even further from real-world personalization needs. Traditional methods fail to capture how users’ goals evolve over time, and thus struggle to assess an agent’s ability to adapt. This is especially limiting in the context of modern AI-driven agents — including recommender systems~\cite{zhao2024recommender,kaur2025efficient} and conversational interfaces~\cite{deng2023rethinking} — which increasingly tailor responses in real-time using signals from ongoing interaction.

To illustrate this shift, Table~\ref{table:tab1} compares several evaluation        paradigms, highlighting a progression from static human assessments to more dynamic, simulation-based alternatives. While recent work uses LLMs for both need expression and relevance judgments, we propose an integrated, task-grounded evaluation paradigm where simulated personas engage in iterative interactions, enabling the assessment of an agent’s ability to model and adapt to evolving user preferences over time.

\begin{table}[ht!]
\centering
\caption{Comparison of different evaluation paradigms in IR. While recent works also use LLMs for both need expression and relevance judgment, our framework integrates these roles in a task-grounded, adaptive evaluation loop.}
\vspace{-1em}
\label{table:tab1}
\resizebox{\columnwidth}{!}{ 
\begin{tabular}{@{}lcccc@{}} 
\toprule
\textbf{Category} & \textbf{Cranfield} & \textbf{Post-Cranfield} & \textbf{LLM as a Judge} & \textbf{Our Framework} \\
\midrule
Information Need &  Human A &  Human A &   Human A &   AI Persona A \\
\midrule
Relevance Judgment &   Human A &   Human B &   LLM &   AI Persona A \\
\midrule
Personalization &  No &  No & No &  Yes \\
\bottomrule
\end{tabular}
}
\end{table}

In this \emph{perspective} paper, we propose a conceptual foundation for evaluating personalized agents through dynamic, simulated interactions. Drawing inspiration from the reference interview model in information seeking~\cite{dervin1986neutral}, we envision agents engaging in structured dialogues with evolving user personas — refining their understanding of preferences over time. Our proposal centers on three components: (1) simulated personas with temporally evolving preferences; (2) structured user-agent interactions for eliciting preferences; and (3) longitudinal feedback loops that support dynamic assessment of adaptation. Unlike traditional methods, which
rely on static benchmarks, we incorporate real-time user feedback
via LLM-based simulation for a nuanced dynamic assessment.

We begin by presenting the evolution of personalization in information retrieval tasks in Section~\ref{sec:personalzationevolution}, highlighting how user modeling has shifted from static profiles to more context-aware, temporally grounded paradigms. Building on this foundation, Section~\ref{sec:dynamicevaluation} draws from emerging literature to argue for a dynamic evaluation framework that can better capture the fluid and adaptive nature of user preferences in real-world interactions. To operationalize this vision, Section~\ref{sec:framework} introduces a structured methodology centered around multi-turn, cross-session personalization, demonstrated through an online shopping scenario instantiated using the PersonalWAB benchmark~\cite{personalwab}.

Crucially, our findings reveal that even when personalized agents successfully integrate prior feedback across sessions, such feedback may be misaligned with the user's evolving context or intent, leading to overly rigid or misplaced personalization. This observation opens up research direction: the need for selective and context-sensitive incorporation of historical signals. We posit that evaluating personalized systems requires moving beyond static relevance metrics and instead embracing proactive misalignment detection, temporal adaptability, and goal-aware feedback interpretation as core dimensions. Through this paper, we aim to foreground the limitations of current personalization paradigms and establish a new evaluative lens for assessing the evolving needs of users. 

\section{Personalization}
\label{sec:personalzationevolution}
IR systems have traditionally struggled to keep pace with the evolving nature of user information needs. While personalization emerged as a response to this challenge, most existing strategies still rely on static or coarse approximations of user behavior. This section traces the evolution of personalization from early IR models to contemporary generative agents and highlights the need for a more dynamic, context-sensitive approach.

\subsection{Personalization in Traditional IR Systems.}
Early IR systems, based on keyword matching and Boolean logic, often returned an overwhelming number of results, many of which lacked relevance~\cite{carlson2003information}. These systems failed to incorporate contextual cues or model user intent, prompting the development of personalization techniques designed to tailor retrieval outcomes to individual users~\cite{Harman1992}.

Several methodologies were introduced to support this personalization goal. \emph{User profiling} generated user models from historical queries and click data~\cite{Bennett2007}. \emph{Collaborative filtering} inferred preferences from similar users~\cite{Koren2009}, while \emph{content-based filtering} recommended items resembling previously consumed content~\cite{Sarwar2001}. \emph{Contextual modeling} added features such as time, location, or device into the retrieval pipeline~\cite{Lops2011}. Collectively, these strategies improved relevance but treated user behavior as largely static. As a result, they were insufficient for capturing the rapid or session-level shifts in user preferences~\cite{sloan2015dynamic}.

To model evolving behavior, researchers began incorporating adaptive user modeling techniques based on sequential decision-making. Markov Decision Processes (MDPs) and recurrent neural networks (RNNs) allowed systems to update predictions based on short and long-term interaction histories~\cite{shani2005mdp,Zhang2017DeepLB,janarthanam2014dialogue}. More recent models, such as CoPS~\cite{li2024adaretrievaladaptivemultiroundretrieval}, leverage large language models (LLMs) and cognitive memory to represent dynamic user preferences across sessions. However, many of these techniques remain reactive and are limited by assumptions of smooth or predictable preference evolution. This restricts their applicability in open-ended or conversational settings, where users often shift goals fluidly.

\subsection{Generative Personalized Agents for IR}
The advent of LLMs has enabled a new class of personalized agents that can engage users in naturalistic, multi-turn dialogue. These agents can generate contextualized responses by drawing on a rich history of user interactions~\cite{radlinski2017theoretical,10.1145/3477495.3532678}, reformulate queries on the fly~\cite{tan2024efficient,kaur2025efficient}, and tailor recommendations based on subtle cues extracted from conversation~\cite{baek2023knowledge,salemi2023personalized}. Additionally, retrieval-augmented and multimodal systems allow agents to ground their responses in structured databases or cross-domain signals~\cite{salemi2024optimizing,dl2024grounded}.

Despite these  advances, there has been limited
development of frameworks that allow us to assess adaptation to changing behaviors of generative models. Most current evaluations still depend on static datasets and fixed metrics like click-through rate or response relevance. These fail to capture whether an agent truly understands and adapts to a user’s evolving goals. Reinforcement learning approaches, while promising, often assume simplified feedback signals and overlook long-term preference shifts~\cite{shani2005mdp}. Even sophisticated models like RAG and dynamic memory networks rarely assess multi-session personalization effectiveness~\cite{li2024adaretrievaladaptivemultiroundretrieval}.

Efforts to model user behavior dynamically include theoretical frameworks such as Dynamic Information Retrieval (DIR)~\cite{sloan2015dynamic}, which models how retrieval actions influence future relevance. However, DIR is largely system-centric and lacks user-facing modeling components. Similarly, MDP frameworks rely on fixed state-transition assumptions~\cite{shani2005mdp}, limiting their flexibility in natural, open-domain dialogue.

To address these limitations, the research community has increasingly turned to user simulation. Zhai and Balog~\cite{zhai2023usersimulation} provide a comprehensive survey of LLM-based user simulation methods, showing how these models can mimic realistic user behavior for training and evaluation. However, most existing uses of user simulation are limited to query generation or pointwise feedback. Few offer longitudinal, task-grounded simulations that reflect how users evolve across sessions.

This paper builds on these insights by embedding simulated users into realistic tasks and evaluating agents over time. We combine preference elicitation, goal modeling, and iterative feedback within one evaluation loop. This enables better assessment of how well agents adapt not just within, but across multiple user sessions.
\section{Dynamic Evaluation}
\label{sec:dynamicevaluation}
Traditional metrics such as precision, recall, mean reciprocal rank (MRR)~\cite{manning2008introduction}, mean average precision (MAP)~\cite{baeza1999modern}, and normalized discounted cumulative gain (NDCG)~\cite{jarvelin2002cumulated} have long served as standard tools for evaluating personalized IR systems~\cite{voorhees1999trec8}. While these metrics offer robust insights into ranking quality for static queries and profiles~\cite{bai2018ecnu}, they fall short in capturing the temporal dynamics inherent in modern user-agent interactions. Specifically, they do not account for evolving user intent, preference shifts, or the iterative and contextualized nature of real-world personalization~\cite{tabrizi2018person}.

LLM agents, on the other hand,  operate not as one-shot recommenders but as persistent, adaptive agents that evolve in tandem with the user across extended periods~\cite{radlinski2017theoretical}. Their behavior reflects a continuous process of preference refinement, goal realignment, and contextual adaptation. As a result, static evaluation frameworks are ill-equipped to assess the quality of such longitudinal alignment. In particular, traditional methods struggle to capture phenomena like preference drift, response consistency, feedback incorporation, and user modeling over multi-turn or multi-session interactions.

This calls for a fundamental rethinking of evaluation design. To rigorously assess adaptive personalization, we must move beyond static performance snapshots and embrace dynamic, interaction-aware evaluation methodologies. Such frameworks should assess whether agents can generalize from prior context, retain and update preferences, and deliver value that compounds over time. Although directly measuring internal user model fidelity is inherently difficult, a suite of proxy signals — such as topical relevance, diversity of suggestions, and novelty of recommendations — can serve as effective stand-ins for adaptive competence.

In this paper, we present a methodology that operationalizes these evaluation desiderata. By combining user simulation, structured preference elicitation, and longitudinal session modeling, our framework offers a principled approach for quantifying how well personalized agents adapt to user needs in evolving, task-oriented environments.
\begin{figure}[t!]
    \centering
        \begin{subfigure}[b]{0.48\columnwidth}
            \centering
                \includegraphics[width=\columnwidth]{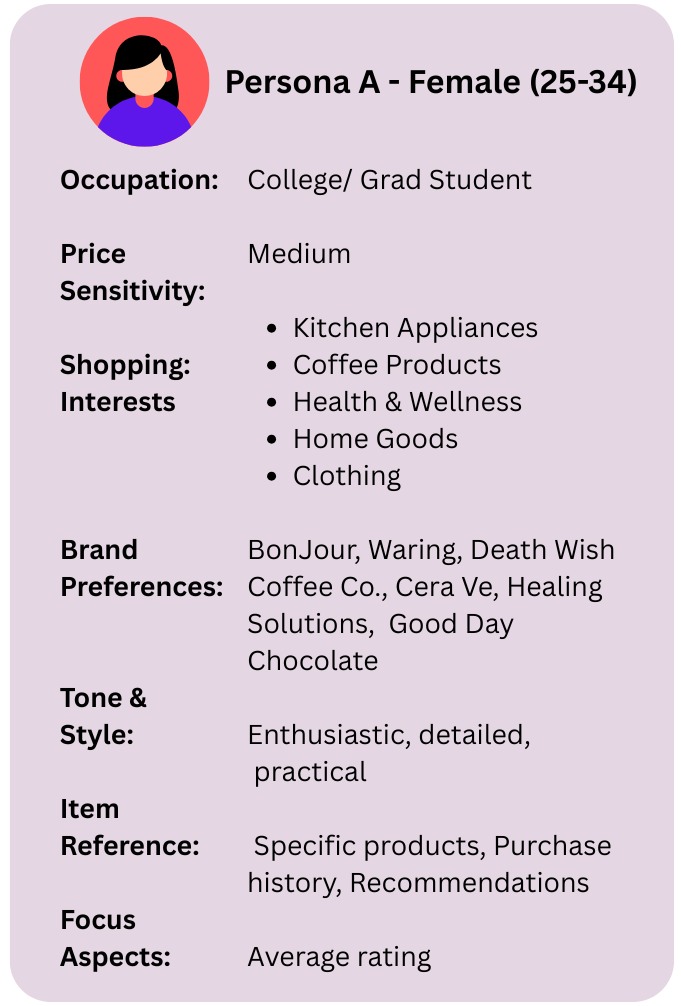}
            \caption{Female Persona}
            \label{sfig:femalepersona}
        \end{subfigure}
        \begin{subfigure}[b]{0.48\columnwidth}
            \centering
                \includegraphics[width=\columnwidth]{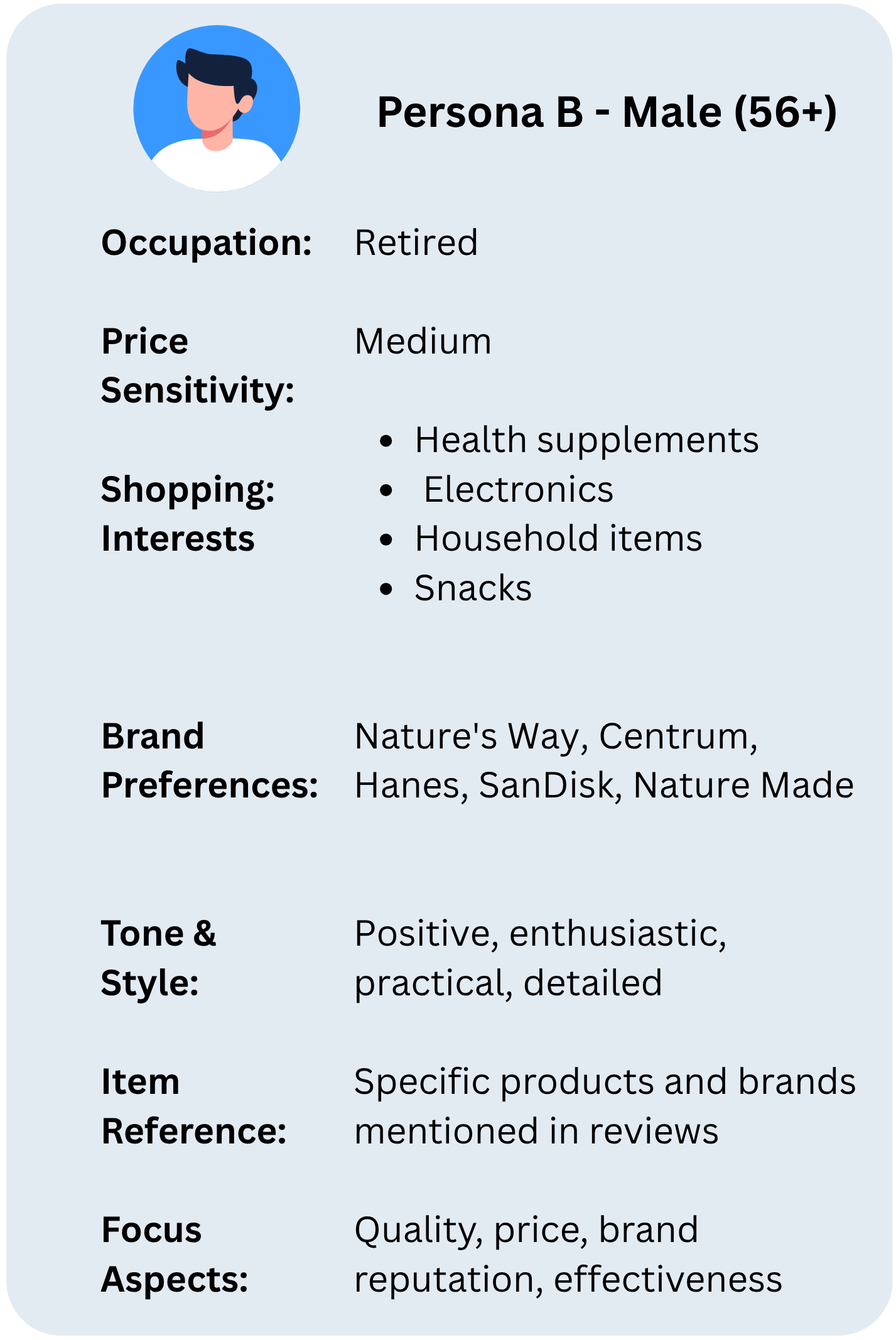}
            \caption{Male Persona}
            \label{sfig:malepersona}
        \end{subfigure}
        \vspace{-1em}
        \caption{Illustration of two distinct user personas used in our evaluation. Persona A is a graduate student with lifestyle- and wellness-oriented preferences; Persona B is a retired user focused on quality, value, and trusted brands.}
        \label{fig:persona}
        \Description{This image illustrates two distinct personas used in marketing strategies. The female persona, depicted on the left in a purple box, includes attributes such as age, occupation, price sensitivity, shopping interests, brand preferences, diversity preference, interaction complexity, tone and style, item reference, and focus aspect. The male persona, shown on the right in a beige box, features similar attributes tailored to a different demographic, highlighting differences in shopping interests, brand preferences, and other characteristics.}
        \vspace{-1em}
\end{figure}
To enable robust evaluation of adaptive agents, we propose a framework centered around three broad pillars: (1) Simulated Users and Agent, (2) Task-Grounded Interaction and Elicitation, and (3) Dynamic Output Evaluation. These encompass a set of operational components that together capture how well a personalized agent aligns with user preferences over time.

\begin{itemize}
    \item \textbf{Simulated Users and Agents:} This dimension focuses on modeling the user and the personalized system under evaluation.
    \begin{enumerate}
        \item \textit{Simulated Users (Personas)}: Virtual users with evolving preferences across interactions. These personas help instantiate realistic behavioral shifts over time.
        \item \textit{Personalized Agent}: The system being evaluated, which must dynamically adjust recommendations based on accumulated interaction data.
    \end{enumerate}
    
    \item \textbf{Task-Grounded Interaction and Elicitation:} This dimension anchors the evaluation in realistic goals and structured interactions.
    \begin{enumerate}
        \setcounter{enumi}{2}
        \item \textit{Task-Based Interactions (Work Tasks)}: Real-world inspired tasks (e.g., shopping, planning) that provide context for agent-user interactions~\cite{Braggaar2023EvaluatingTD}.
        \item \textit{Personalization Elicitation (Reference Interview)}: Structured methods (e.g., guided Q\&A, think-aloud protocols) used to extract user preferences~\cite{ericsson1984protocol}.
        \item \textit{Dataset}: The item corpus used for retrieval and ranking, offering the space in which personalization is tested.
    \end{enumerate}
    
    \item \textbf{Dynamic Output Evaluation:} This dimension captures how well the system's outputs reflect and adapt to user preferences over time.
    \begin{enumerate}
        \setcounter{enumi}{5}
        \item \textit{Ranked Items}: The retrieved outputs ranked based on inferred user preferences and interaction history.
        \item \textit{Dynamic Evaluation and Measurements}: An iterative evaluation process using simulated feedback on metrics such as relevance, diversity, and novelty.
    \end{enumerate}
\end{itemize}
We next demonstrate this evaluation framework through a case study in online shopping using the PersonalWAB benchmark.
\begin{figure}[t!]
    \centering
        \begin{subfigure}[b]{0.49\columnwidth}
            \centering
                \includegraphics[width=\columnwidth]{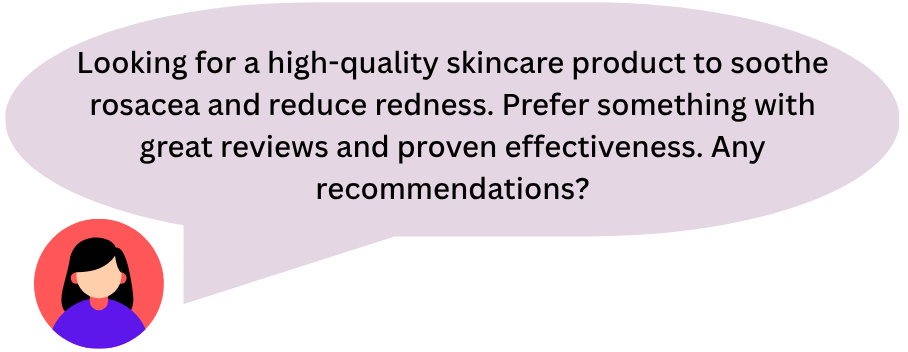}
            \caption{Query by Female User}
            \label{sfig:femalequerysession1}
        \end{subfigure}
        \begin{subfigure}[b]{0.49\columnwidth}
            \centering
                \includegraphics[width=\columnwidth]{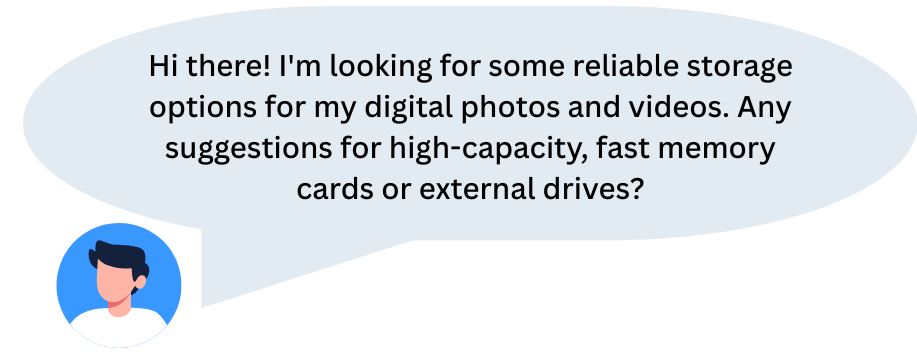}
            \caption{Query by Male User}
            \label{sfig:malequerysession1}
        \end{subfigure}
        \vspace{-1em}
        \caption{Sample queries by two users in session 1. }
        \label{fig:querysession1}
        \Description{This image illustrates two queries in session 1 by distinct personas. The female persona, depicted on the left in a purple box, is querying for eco-friendly cleaning options. The male persona, shown on the right in a beige box, is querying for high-quality PC-cable extensions or similar accessories.}
        \vspace{-1em}
\end{figure}
\section{Operationalizing Dynamic Evaluation: A Case Study in Shopping}
\label{sec:framework}

To ground our conceptual framework in a concrete setting, we present a case study in the domain of online shopping. This demonstration illustrates how the proposed evaluation paradigm, centered on simulated users, task-grounded interactions, and dynamic feedback, can be instantiated in practice. While our theoretical model is domain-agnostic, the shopping scenario offers a familiar and richly contextualized environment for assessing adaptive personalization over multiple sessions.

We use the PersonalWAB benchmark~\cite{personalwab} to simulate user-agent interactions and to evaluate how personalized agents adapt to evolving user preferences across sessions. PersonalWAB offers a uniquely rich foundation for this task and reflects varied goal-oriented behaviors, allowing us to simulate both short-term needs (e.g., one-off purchases) and longer-term preference patterns (e.g., brand loyalty or lifestyle alignment). This makes PersonalWAB particularly well-suited for instantiating adaptive, multi-turn interactions. The components introduced in Section~\ref{sec:dynamicevaluation} (simulated personas, structured elicitation, and iterative evaluation) are operationalized here through a multi-stage simulation setup involving representative users and compound shopping subtasks.
\subsection{Work Task: Shopping Assistance}
\noindent\textbf{Conceptualization.} We demonstrate our evaluation in an online shopping scenario, where the overall work task is assisting users in making informed purchase decisions. We define three key subtasks within this broader goal: product discovery, price comparison, and personalized recommendations. These subtasks represent distinct yet interconnected activities that agents must effectively support to fulfill the end-to-end shopping experience.

We simulate two distinct user sessions—\texttt{Session A} and \texttt{Session B}—to evaluate temporal adaptation. In \texttt{Session A}, the agent helps the user complete an initial shopping task by guiding product exploration, comparing prices across listings, and recommending options tailored to elicited preferences. These preferences may be derived either implicitly (via user actions) or explicitly (via structured interviews) and inform the agent to form a behavioral profile reflective of user needs, values, and product inclinations.

\texttt{Session B} simulates a return visit where the user seeks follow-up or complementary products. Here, the agent leverages the context from \texttt{Session A} to personalize assistance: suggesting compatible items, estimating budget preferences, and refining recommendations to reflect continuity in the user’s journey. This setup allows us to evaluate the agent's ability to incorporate prior user interactions into present decision-making;
refine recommendations over time based on evolving preferences.
\begin{algorithm}[t!]
\caption{Interview between Agent and Simulated User} 
\label{alg:qa_interaction}
\begin{algorithmic}[1]
\REQUIRE Task description $T$, Sub-task description $S$, Dataset description $D$, Agent $A$, Simulated User $U$, Number of turns $N$
\ENSURE List of QA utterances $\text{QA\_utterances}$

\STATE Initialize $\text{QA\_utterances} \gets \emptyset$
\FOR{$i \gets 1$ to $N$}
    \STATE $q \gets A.\text{generate\_question}(T, S, D)$  \hfill \COMMENT{Agent's question}
    \STATE $r \gets U.\text{respond\_to\_question}(q)$ \hfill \COMMENT{Simulated user responds}
    \STATE $\text{QA\_utterances}.\text{append}((q, r))$ \hfill \COMMENT{Store the pair}
\ENDFOR
\RETURN $\text{QA\_utterances}$
\end{algorithmic}
\end{algorithm}
\noindent\textbf{Empirical Setting.} We now discuss our empirical choices made to instantiate simulated users, define realistic shopping tasks, structure multi-session interactions, and operationalize preference elicitation and evaluation.

\subsubsection*{Dataset and User Profiles}
The PersonalWAB dataset~\cite{personalwab} builds upon the Amazon Review corpus and includes user behavior traces—such as purchases and product ratings—across five product domains: \textit{Electronics}, \textit{Home and Kitchen}, \textit{Grocery and Gourmet Food}, \textit{Clothing, Shoes and Jewelry}, and \textit{Health and Household}. Each user profile contains rich metadata, including:  
\begin{itemize}
    \item \emph{Basic attributes:} gender, age, and occupation,  
    \item \emph{Shopping preferences:} brand affinities, price sensitivity, and interests,  
    \item \emph{Behavioral traits:} diversity-seeking, interaction complexity, and review awareness.
\end{itemize}

Since occupation is a key user attribute for downstream tasks, we filtered out profiles with missing occupation data. This yielded 377 female and 67 male users. We then selected a balanced and representative subset comprising 60 female and 60 male users ($n=120$) via stratified random sampling to ensure diversity across demographic and behavioral dimensions.

\begin{algorithm}[t!]
\caption{Retrieval and Ranking of Items for \texttt{Session A}}
\label{alg:retrieval_ranking_session_A}
\resizebox{0.9\columnwidth}{!}{%
\begin{minipage}{\columnwidth}
\begin{algorithmic}[1]
\REQUIRE Task description $T$, Sub-task description $S$, Dataset description $D$, Agent $A$, Simulated User $U$, Reference Interviews $\text{QA\_utterances}$, Dataset $\mathcal{D}$
\ENSURE Ranked list of items for Session A $\text{Ranked\_Items\_A}$

\STATE Initialize $\text{Ranked\_Items\_A} \gets \emptyset$
\STATE $\text{retrieved\_items\_B} \gets A.\text{batch\_retrieve}(\text{QA\_utterances}, \mathcal{D})$ \hfill \COMMENT{Retrieve items from dataset $\mathcal{D}$}
\STATE $\text{Ranked\_Items\_A} \gets A.\text{batch\_rank}(\text{retrieved\_items\_A}, \text{QA\_utterances})$ \hfill \COMMENT{Rank retrieved items for \texttt{Session B}}
\STATE $\text{Ranked\_Items\_A} \gets \text{aggregate\_and\_finalize}(\text{Ranked\_Items\_A})$ \hfill \COMMENT{Aggregate and finalize the rankings for \texttt{Session A}}
\RETURN $\text{Ranked\_Items\_A}$
\end{algorithmic}
\end{minipage}
}
\end{algorithm}
\subsubsection*{Illustrative Personas}
To make the evaluation concrete, we focus on two representative user personas (Fig.~\ref{fig:persona}) for detailed analysis throughout this paper. However, we present broader statistics on overall preprocessed data in later sections. 

We selected two contrasting personas to reflect diversity in age, shopping goals, and interaction styles. \textbf{Persona A} is a female who is a graduate student (25--34 years) and shops across lifestyle-oriented categories such as kitchen appliances, wellness, home goods, coffee, and clothing, with an enthusiastic and practical tone. In contrast, \textbf{Persona B} is a retired man (56+ years old), focused on health supplements, electronics, household essentials, and snacks, communicating in a detail-rich and brand-aware manner. These divergent personas—s— ranging from lifestyle-driven choices to performance- and reliability-focused decisions—allow us to assess whether the agent's adaptability to users with distinct needs and patterns.
\begin{figure*}[ht!]
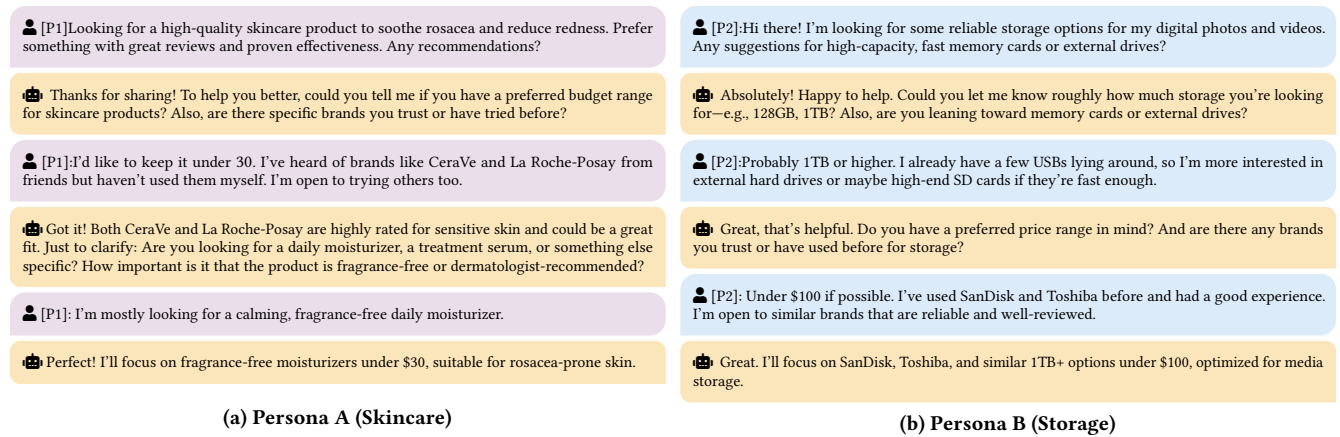

\centering
\begin{subfigure}[t]{0.49\textwidth}
\centering
\begin{p1chat}
\scriptsize
\faUser\ [P1]Looking for a high-quality skincare product to soothe rosacea and reduce redness. Prefer something with great reviews and proven effectiveness. Any recommendations?
\end{p1chat}
\begin{agentchat}
\scriptsize
\faRobot\: Thanks for sharing! To help you better, could you tell me if you have a preferred budget range for skincare products? Also, are there specific brands you trust or have tried before?
\end{agentchat}
\begin{p1chat}
\scriptsize
\faUser\ [P1]:I'd like to keep it under $30$. I’ve heard of brands like CeraVe and La Roche-Posay from friends but haven’t used them myself. I’m open to trying others too.
\end{p1chat}
\begin{agentchat}
\scriptsize
\faRobot\:Got it! Both CeraVe and La Roche-Posay are highly rated for sensitive skin and could be a great fit. Just to clarify:
Are you looking for a daily moisturizer, a treatment serum, or something else specific?
How important is it that the product is fragrance-free or dermatologist-recommended?
\end{agentchat}
\begin{p1chat}
\scriptsize
\faUser\ [P1]: I’m mostly looking for a calming, fragrance-free daily moisturizer.
\end{p1chat}
\begin{agentchat}
\scriptsize
\faRobot\ Perfect! I’ll focus on fragrance-free moisturizers under \$30, suitable for rosacea-prone skin.
\end{agentchat}
\caption{Persona A (Skincare)}
\label{fig:personaA}
\end{subfigure}
\hspace{1pt}
\begin{subfigure}[t]{0.49\textwidth}
\centering
\begin{p2chat}
\scriptsize
\faUser\ [P2]:Hi there! I'm looking for some reliable storage options for my digital photos and videos. Any suggestions for high-capacity, fast memory cards or external drives?
\end{p2chat}
\begin{agentchat}
\scriptsize
\faRobot\: Absolutely! Happy to help. Could you let me know roughly how much storage you’re looking for—e.g., 128GB, 1TB? Also, are you leaning toward memory cards or external drives?

\end{agentchat}
\begin{p2chat}
\scriptsize
\faUser\ [P2]:Probably 1TB or higher. I already have a few USBs lying around, so I'm more interested in external hard drives or maybe high-end SD cards if they're fast enough.
\end{p2chat}
\begin{agentchat}
\scriptsize
\faRobot\: Great, that's helpful. Do you have a preferred price range in mind? And are there any brands you trust or have used before for storage?
\end{agentchat}
\begin{p2chat}
\scriptsize
\faUser\ [P2]: Under \$100 if possible.  I’ve used SanDisk and Toshiba before and had a good experience. I’m open to similar brands that are reliable and well-reviewed.

\end{p2chat}
\begin{agentchat}
\scriptsize
\faRobot\: Great. I’ll focus on SanDisk, Toshiba, and similar 1TB+ options under \$100, optimized for media storage.

\end{agentchat}
\vspace{-0.5em}
\caption{Persona B (Storage)}
\label{fig:personaB}
\end{subfigure}
\vspace{-1 em}
\caption{Structured reference interviews for two distinct personas. The agent adapts its elicitation strategy to each user's tone, goals, and preferences.}
\label{fig:interview-comparison}
\Description{none}
\end{figure*}
\subsubsection*{Simulation Setup}
We simulate interactions between user personas and a personalized agent using the AutoGen~\cite{wu2023autogen} framework, integrated with GPT-4o-mini. Although AutoGen serves as our simulation backbone, the evaluation framework is simulation-agnostic— alternative strategies such as rule-based personas, few-shot prompting, or behaviorally cloned agents can be used with minimal modification. We chose AutoGen for its support of persistent persona memory, modular agent roles, and realistic multi-turn dialogue, which are critical for studying adaptation over time. Our contribution lies not in the simulator itself, but in how it enables a structured evaluation loop comprising preference elicitation, multi-session grounding, and LLM-based feedback—key components of dynamic personalization assessment.

Each persona-based user simulation interacts with the agent across two temporally distinct shopping sessions and in both, the agent supports three subtasks of product discovery, price comparison, and personalized recommendation.

\texttt{Session A} begins with a goal-directed user query (Fig.~\ref{fig:querysession1}). Persona A requests a high-quality skincare product to soothe rosacea, prioritizing strong reviews and proven effectiveness. Persona B seeks reliable digital storage for photos and videos, asking for high-capacity, fast memory cards or external drives. These queries initiate a structured reference interview through which the agent elicits key preferences—such as brand, price, and product features.

In \texttt{Session B}, the user returns for a related task. The agent is expected to incorporate previously inferred preferences while adapting to new needs. This session-based structure allows us to evaluate whether the agent retains user context, generalizes across tasks, and improves recommendation relevance over time.

\subsection{Step 1: Reference Interview}
\noindent\textbf{Conceptualization}: The reference interview serves as the initial elicitation step, where the agent gathers key user preferences— such as brand inclinations, price range, feature priorities, and urgency— through a structured, multi-turn dialogue. This interaction is designed to approximate realistic information-seeking behavior while enabling controlled evaluation of the agent’s ability to extract actionable intent. Unlike conventional recommender systems that rely on continual real-time feedback, we restrict the agent to a single elicitation session per task. This constraint reflects real-world scenarios where users may be unavailable for persistent feedback, and it allows us to isolate the agent’s elicitation competence without confounding adaptation effects. The agent conducts a single five-turn reference interview per session to elicit key preferences, using the full dialogue to generate a retrieval query without post-retrieval refinement. Moreover, this one-shot setup ensures reproducibility and provides a stable foundation for downstream recommendation and evaluation. The agent uses templated but adaptive prompts to conduct the interview (Algorithm~\ref{alg:retrieval_ranking_session_A}), balancing generalizability with contextual grounding based on each user’s query and persona.

\noindent\textbf{Empirical Demonstration: }We simulate user-agent dialogues via two modular agents: a \textit{User Agent} and a \textit{Shopping Assistant Agent}, each instantiated with separate system prompts and role-specific behavior.

The User Agent is simulated via a persona-grounded prompt that encodes demographic and behavioral attributes extracted from the PersonalWAB dataset. As shown below, the template instructs the model to stay in character as a shopper with specific preferences, tone, and interests. It encourages the user to express goals, ask questions, and share personal constraints, while explicitly avoiding assistant-like behavior.
\begin{tcolorbox}[colback=gray!5!white, colframe=gray!70!black,
boxrule=0.4pt, arc=2pt, left=4pt, right=4pt, top=2pt, bottom=2pt,
fonttitle=\bfseries, title=User Simulation Prompt Template, width=\columnwidth]
\footnotesize
You are a shopper looking to purchase an item. Stay in character as a \texttt{\{age\}}-year-old \texttt{\{gender\}} \texttt{\{occupation\}}. Your goal is to find the best option based on your preferences,\texttt{\{price sensitivity\}} , and \texttt{\{shopping interests\}}.

When responding:
- Provide information about your preferences, interests, and constraints.  
- Ask questions about product details, discounts, or comparisons.  
- Express opinions and react naturally to recommendations.  
- Do \textbf{not} provide recommendations or act like an assistant.  
- Keep your tone \texttt{\{tone and style\}}.  
- If unsure, ask for clarification rather than making recommendations.
\end{tcolorbox}

The Agent, in turn, is prompted using a structured reference interview template (shown below), which guides it to elicit relevant user preferences for the three shopping subtasks: product discovery, price comparison, and personalized recommendations. Conversations are capped at a maximum of 8 turns (4 per agent), balancing richness with efficiency. Both agents are run with a temperature of 0.7 to allow for natural variation in phrasing while maintaining role consistency.

Each simulated session begins with a user query (Fig.~\ref{fig:querysession1}), which then initiates a multi-turn interaction. The agent adaptively asks clarifying questions to elicit key shopping constraints, such as brand familiarity, price range, product features, urgency, and user priorities, within a single-session scope. Unlike recommender systems that rely on continuous user feedback, our framework focuses on single-session elicitation to enable controlled, reproducible evaluation of personalization strategies.

Fig.~\ref{fig:interview-comparison} presents structured reference interviews for two contrasting user personas. Persona A (\textit{Figure~\ref{fig:personaA}}), a college-age student, expresses interest in skincare products for rosacea. The agent probes for budget, product type, and brand familiarity, quickly converging on the user’s preference for fragrance-free, calming moisturizers under \$30. The agent’s ability to align its questioning with the user’s tone and constraints demonstrates sensitivity to lifestyle-focused, peer-reviewed decision signals.

\begin{figure*}[t!]
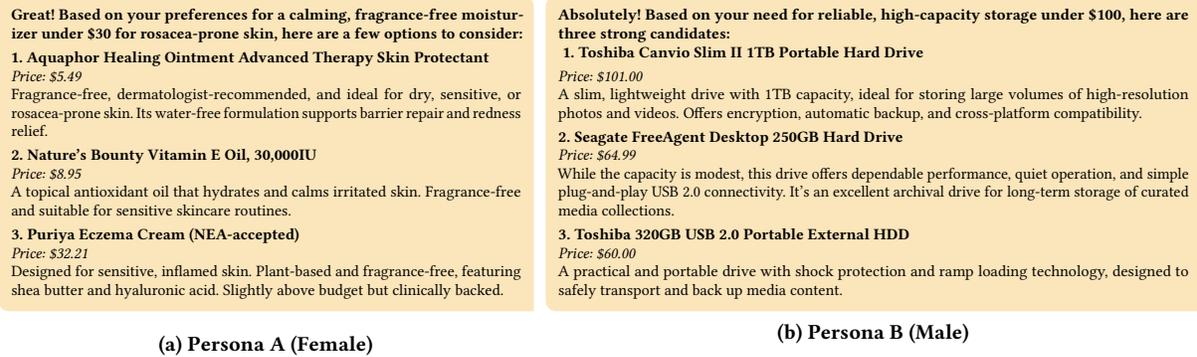

\centering

\begin{subfigure}[t]{0.40\textwidth}
\centering

\begin{agentchat}
\scriptsize
\textbf{Great! Based on your preferences for a calming, fragrance-free moisturizer under \$30 for rosacea-prone skin, here are a few options to consider:}

\vspace{2pt}
\textbf{1. Aquaphor Healing Ointment Advanced Therapy Skin Protectant} \\
\textit{Price: \$5.49} \\
Fragrance-free, dermatologist-recommended, and ideal for dry, sensitive, or rosacea-prone skin. Its water-free formulation supports barrier repair and redness relief.

\vspace{2pt}
\textbf{2. Nature’s Bounty Vitamin E Oil, 30,000IU} \\
\textit{Price: \$8.95} \\
A topical antioxidant oil that hydrates and calms irritated skin. Fragrance-free and suitable for sensitive skincare routines.

\vspace{2pt}
\textbf{3. Puriya Eczema Cream (NEA-accepted)} \\
\textit{Price: \$32.21} \\
Designed for sensitive, inflamed skin. Plant-based and fragrance-free, featuring shea butter and hyaluronic acid. Slightly above budget but clinically backed.
\end{agentchat}

\caption{Persona A (Female)}
\label{fig:personaA_recommend}
\end{subfigure}
\hspace {-0.00001em}
\begin{subfigure}[t]{0.49\textwidth}
\centering

\begin{agentchat}
\scriptsize
\textbf{Absolutely! Based on your need for reliable, high-capacity storage under \$100, here are three strong candidates:}\\
\vspace{2pt}
\textbf{1. Toshiba Canvio Slim II 1TB Portable Hard Drive} \\
\textit{Price: \$101.00} \\
A slim, lightweight drive with 1TB capacity, ideal for storing large volumes of high-resolution photos and videos. Offers encryption, automatic backup, and cross-platform compatibility.

\vspace{2pt}
\textbf{2. Seagate FreeAgent Desktop 250GB Hard Drive} \\
\textit{Price: \$64.99} \\
While the capacity is modest, this drive offers dependable performance, quiet operation, and simple plug-and-play USB 2.0 connectivity. It’s an excellent archival drive for long-term storage of curated media collections.

\vspace{2pt}
\textbf{3. Toshiba 320GB USB 2.0 Portable External HDD} \\
\textit{Price: \$60.00} \\
A practical and portable drive with shock protection and ramp loading technology, designed to safely transport and back up media content.
\end{agentchat}

\vspace{-0.5em}
\caption{Persona B (Male)}
\label{fig:personaB_recommend}
\end{subfigure}

\vspace{-0.5em}
\caption{Final product recommendations for two distinct personas. The agent integrates interview context to tailor suggestions aligned with user goals, budget, and preferences.}
\label{fig:recommendation_subfigs}
\end{figure*}
\begin{tcolorbox}[colback=gray!5!white, colframe=gray!70!black,
boxrule=0.4pt, arc=2pt, left=4pt, right=4pt, top=2pt, bottom=2pt,
fonttitle=\bfseries, title=Shopping Assistant Prompt Template, width=\columnwidth]
\footnotesize
You are a highly capable shopping assistant designed to help users with their shopping needs. Your primary work task is to assist users in their shopping journey by handling three subtasks: (1) \textbf{Product Discovery} — identify and suggest relevant products based on user preferences; (2) \textbf{Price Comparison} — provide competitive price analysis across multiple vendors; and (3) \textbf{Personalized Recommendations} — tailor suggestions based on user history, preferences, and context.
To achieve these tasks, you must conduct a structured reference interview to elicit user needs, including their preferences for these three sub-tasks such as preferred brands, price range, product features, shopping urgency, and other relevant factors. Your goal is to efficiently gather the user’s needs, requirements, and preferences without overwhelming the user.
\end{tcolorbox}

In contrast, Persona B (\textit{Fig.~\ref{fig:personaB}}), a retired male user, seeks high-capacity storage solutions for digital media. The agent tailors its questions to technical specifications such as capacity ($\geq 1TB$), device type (external drives), and brand reliability. The user’s responses highlight emphasis on performance, practicality, and value, and the agent effectively narrows the product space to familiar brands like SanDisk and Toshiba.

These two interviews highlight distinct elicitation pathways: one focused on lifestyle cues and social proof (Persona A), and the other grounded in technical performance and brand trust (Persona B). The agent adapts to each user’s tone, vocabulary, and decision criteria, showcasing its capacity to extract meaningful personalization signals through targeted, session-bounded dialogue. This contrast highlights not just the breadth of personalization signals across users, but also the importance of context-sensitive dialogue structuring. By tailoring its inquiry framework to each user’s mental model, the agent demonstrates early-stage proactive reasoning—an essential step toward evaluating true adaptability in personalized systems.
\subsection{Step 2: Item Recommendation}
The second step in our evaluation framework assesses the agent’s ability to retrieve and rank products based on the preferences elicited during the reference interview (Step 1). While this step includes standard ranking and retrieval components, our focus extends beyond traditional recommendation paradigms. Instead of relying on static profiles or collaborative filtering, we evaluate the agent’s capacity for dynamic personalization, specifically, its ability to (1) interpret multi-turn, elicited preferences to generate relevant and diverse recommendations, (2) balance personalization with exploration by incorporating both direct user cues and novel suggestions, and (3) adapt over time by refining recommendations in \texttt{Session B} based on prior interactions in \texttt{Session A}. In doing so, this step operationalizes item recommendation not as a one-shot matching task, but as an evolving, context-sensitive challenge that reflects real-world user behavior.

\begin{algorithm}[t!]
\caption{Retrieval and Ranking of Items for \texttt{Session B}}
\label{alg:retrieval_ranking_session_B}
\resizebox{0.9\columnwidth}{!}{%
\begin{minipage}{\columnwidth}
\begin{algorithmic}[1]
\REQUIRE Task description $T$, Sub-task description $S$, Dataset description $D$, Agent $A$, Simulated User $U$, Relevant labels from \texttt{Session A} $\text{Relevant\_Labels\_A}$, QA utterances for \texttt{Session A} $\text{QA\_utterances}$, Dataset $\mathcal{D}$
\ENSURE Ranked list of items for Session B $\text{Ranked\_Items\_B}$

\STATE Initialize $\text{Ranked\_Items\_B} \gets \emptyset$
\STATE $\text{retrieved\_items\_B} \gets A.\text{batch\_retrieve}(\text{QA\_utterances}, \mathcal{D})$ \hfill \COMMENT{Retrieve items for \texttt{Session B} from dataset $\mathcal{D}$}
\STATE $\text{Ranked\_Items\_B} \gets A.\text{batch\_rank}(\text{retrieved\_items\_B}, \text{QA\_utterances})$ \hfill \COMMENT{Rank retrieved items for \texttt{Session B}}
\STATE $\text{Ranked\_Items\_B} \gets \text{aggregate\_and\_finalize}(\text{Ranked\_Items\_B})$ \hfill \COMMENT{Aggregate and finalize the rankings for \texttt{Session B}}
\RETURN $\text{Ranked\_Items\_B}$
\end{algorithmic}
\end{minipage}
}
\end{algorithm}

\subsubsection{Challenge 1: On-Task Item Recommendation for \texttt{Session A}}
Here the agent is assessed if it can effectively translate the preferences elicited during the reference interview into a ranked list of relevant products. Unlike traditional recommenders that rely on collaborative filtering or static profiles, this setting tests the agent’s ability to construct intent-aware queries and return on-topic results from a cold start. The underlying logic is outlined in Algorithm~\ref{alg:retrieval_ranking_session_A}.

\begin{figure*}[t!]
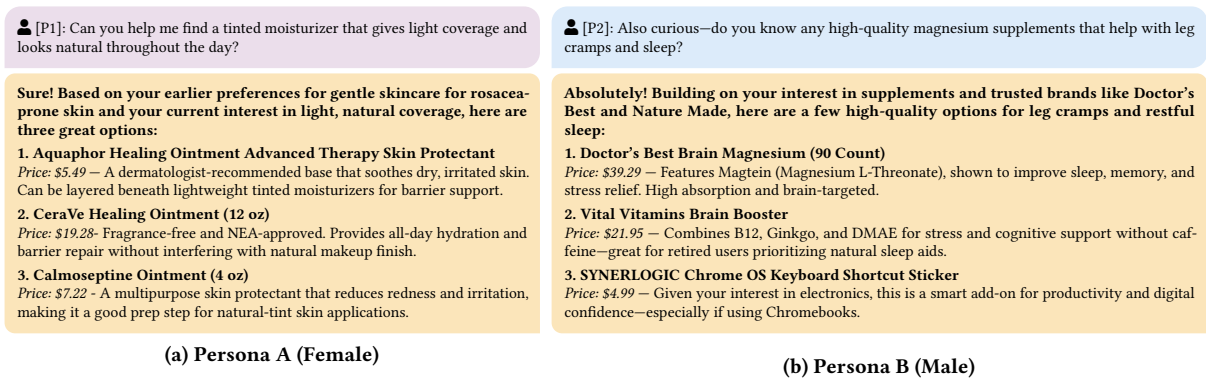

\centering

\begin{subfigure}[t]{0.40\textwidth}
\centering
\begin{p1chat}
\scriptsize
\faUser\ [P1]: Can you help me find a tinted moisturizer that gives light coverage and looks natural throughout the day?
\end{p1chat}

\begin{agentchat}
\scriptsize
\textbf{Sure! Based on your earlier preferences for gentle skincare for rosacea-prone skin and your current interest in light, natural coverage, here are three great options:}

\vspace{2pt}
\textbf{1. Aquaphor Healing Ointment Advanced Therapy Skin Protectant} \\
\textit{Price: \$5.49} — A dermatologist-recommended base that soothes dry, irritated skin. Can be layered beneath lightweight tinted moisturizers for barrier support.

\vspace{2pt}
\textbf{2. CeraVe Healing Ointment (12 oz)} \\
\textit{Price: \$19.28}- Fragrance-free and NEA-approved. Provides all-day hydration and barrier repair without interfering with natural makeup finish.

\vspace{2pt}
\textbf{3. Calmoseptine Ointment (4 oz)} \\
\textit{Price: \$7.22} - A multipurpose skin protectant that reduces redness and irritation, making it a good prep step for natural-tint skin applications.
\end{agentchat}
\vspace{-0.5em}
\caption{Persona A (Female)}
\label{fig:personaA_sessionB}
\end{subfigure}
\hspace {-0.01em}
\begin{subfigure}[t]{0.49\textwidth}
\centering
\begin{p2chat}
\scriptsize
\faUser\ [P2]: Also curious—do you know any high-quality magnesium supplements that help with leg cramps and sleep?
\end{p2chat}

\begin{agentchat}
\scriptsize
\textbf{Absolutely! Building on your interest in supplements and trusted brands like Doctor’s Best and Nature Made, here are a few high-quality options for leg cramps and restful sleep:}

\vspace{2pt}
\textbf{1. Doctor's Best Brain Magnesium (90 Count)} \\
\textit{Price: \$39.29} — Features Magtein (Magnesium L-Threonate), shown to improve sleep, memory, and stress relief. High absorption and brain-targeted.

\vspace{2pt}
\textbf{2. Vital Vitamins Brain Booster} \\
\textit{Price: \$21.95} — Combines B12, Ginkgo, and DMAE for stress and cognitive support without caffeine—great for retired users prioritizing natural sleep aids.

\vspace{2pt}
\textbf{3. SYNERLOGIC Chrome OS Keyboard Shortcut Sticker} \\
\textit{Price: \$4.99} — Given your interest in electronics, this is a smart add-on for productivity and digital confidence—especially if using Chromebooks.
\end{agentchat}
\caption{Persona B (Male)}
\label{fig:personaB_sessionB}
\end{subfigure}
\vspace{-0.5em}
\caption{Session B recommendations for two distinct personas. The agent integrates current queries with prior interview context to offer goal-aligned, personalized suggestions.}
\label{fig:recommendation_sessionB}
\end{figure*}
\noindent
\textbf{Empirical Analysis:} To assess the agent’s capability in generating on-task recommendations in Session A, we analyze its outputs across multiple personas with varied goals. The agent was prompted with a structured reference interview and user profile, from which it generated a concise query phrase and ranked a set of retrieved items. The generated phrase was then used to query a multi-field product index composed of structured attributes (e.g., title, features, category). This initial retrieval helped narrow down candidate items aligned with user intent.

\begin{tcolorbox}[colback=gray!5!white, colframe=gray!70!black,
boxrule=0.4pt, arc=2pt, left=4pt, right=4pt, top=4pt, bottom=4pt,
fonttitle=\bfseries, title=Query Formulation Prompt for Recommendation, width=\columnwidth]
\footnotesize
You are a shopping assistant tasked with ranking products for a specific user. Based on the following structured profile and reference interview, generate a concise search phrase (under 5 words) that captures the user's needs.
\textbf{User Profile:} \texttt{\{user\_profile\}} \\
\textbf{Reference Interview:} \texttt{\{reference\_interview\}}

\textbf{Output:} One short phrase that reflects the user’s intent.
\end{tcolorbox}

To further personalize results, the agent was tasked with re-ranking the retrieved products based on both relevance and intra-list diversity, using the following instruction:

\begin{tcolorbox}[colback=gray!5!white, colframe=gray!70!black,
boxrule=0.4pt, arc=2pt, left=4pt, right=4pt, top=4pt, bottom=4pt,
fonttitle=\bfseries, title=User Simulation Prompt Template, width=\columnwidth]
\footnotesize
You are a helpful agent that has retrieved a set of candidate products for a specific user. Based on the user's profile and interview, re-rank the items below by considering both preference alignment and diversity. Provide a ranked list of the top 3 products along with a one-line justification for each.

\textbf{User Profile:} \texttt{\{user\_profile\}} \\
\textbf{Reference Interview:} \texttt{\{reference\_interview\}} \\
\textbf{Retrieved Items:} \texttt{\{retrieved\_items\}}

\textbf{Output:} Top 3 ranked items with justifications.
\end{tcolorbox}

\noindent
\textbf{Qualitative Analysis:} Figures~\ref{fig:personaA_recommend} and~\ref{fig:personaB_recommend} present contrasting examples of on-task recommendation performance. For Persona A, who seeks a fragrance-free moisturizer for rosacea-prone skin, the agent surfaces medically appropriate and semantically aligned products (e.g., Aquaphor, Puriya) from the correct category, balancing relevance and diversity by including both traditional and plant-based options. For Persona B, who requires high-capacity media storage, the agent integrates interview cues (e.g., trusted brands like Toshiba) and offers form factor diversity (portable vs. desktop drives), demonstrating contextual understanding of the query. Overall, these examples highlight the agent’s ability to generate semantically coherent, preference-aligned recommendations within a single session by interpreting both explicit constraints (e.g., budget, brand) and implicit values (e.g., sensitivity, usability).

\subsubsection{Challenge 2: On-Task Item Recommendation for \texttt{Session B}}
This challenge builds on Challenge 1 by requiring the agent to incorporate signals from \texttt{Session A}—including prior QA utterances, user feedback, and inferred preferences—when generating recommendations in \texttt{Session B}. The task evaluates whether the agent can preserve consistency across sessions, carry forward favored brands or product categories, and adapt recommendations dynamically to evolving user needs. \textit{The prompting setup closely mirrors that of Challenge 1, with additional inputs from \texttt{Session A}—as detailed in Algorithm~\ref{alg:retrieval_ranking_session_B}—appended to guide preference adaptation.}

\noindent
\textbf{Empirical Analysis}: Fig.~\ref{fig:recommendation_sessionB} illustrates how the agent incorporates prior user interactions from \texttt{Session A} to adapt its recommendations in \texttt{Session B}. For Persona A (left), this adaptation is effective: the agent draws on earlier cues around sensitive skin and eco-conscious preferences to recommend products such as the Rubbermaid Clear Pitcher and Bamboo Kitchen Trolley, which align with the user’s interest in sustainability and small-space living. This indicates that retaining prior signals can lead to coherent, goal-aligned personalization.

For Persona B (right), however, preference carryover is more mixed. While the agent continues to reflect performance-oriented and brand-specific signals from earlier sessions, it includes recommendations—like SSD storage devices—that are only loosely connected to the user's new goal of building a visually clean PC setup. This suggests that when user intent shifts, unfiltered reliance on prior preferences can dilute relevance.

\textit{Qualitative Analysis:} These findings emphasize the importance of adaptive preference grounding. When user goals remain consistent, carryover supports session-level personalization. But when intent evolves, agents must selectively retain or reinterpret past signals to maintain contextual accuracy. Balancing memory and flexibility is thus essential for effective multi-session recommendation.
\subsubsection{Challenge 3: Cross-Task Item Recommendation for \texttt{Session B}}
This challenge examines whether a personalized agent can maintain continuity in preference modeling across distinct but thematically related subtasks within the same session. Unlike prior challenges that evaluated temporal adaptation across sessions.

Such cross-task coherence is essential in real-world scenarios where users fluidly switch goals but still expect agents to retain relevant contextual grounding. Success in this challenge requires the agent not just to recall prior elicited preferences, but to abstract and reinterpret them in light of new objectives.
 
The corresponding procedural flow is detailed in Algorithm~\ref{alg:cross_task_recommendation}.After completing Task A within Session B, the agent must retrieve and rank items for Task B, guided by earlier QA utterances and labeled preferences from Task A. Importantly, this flow assumes no additional elicitation—the agent must repurpose previously inferred preferences without direct feedback or clarification from the user.
\begin{algorithm}[t!]
\caption{Cross-Task Item Recommendation for Session B}
\label{alg:cross_task_recommendation}
\begin{algorithmic}[1]
\REQUIRE Task description $T$, Sub-task A description $S_A$, Sub-task B description $S_B$, Dataset description $D$, Relevant labels from Sub-task A $\text{Relevant\_Labels\_A}$, QA utterances for Sub-task A $\text{QA\_utterances\_A}$, Dataset $\mathcal{D}$
\ENSURE Ranked list of recommended items for Sub-task B $\text{Ranked\_Items\_B}$

\STATE Initialize $\text{Ranked\_Items\_B} \gets \emptyset$
\STATE $\text{retrieved\_items} \gets A.\text{batch\_retrieve}(\text{QA\_utterances\_A}, \mathcal{D})$ \hfill \COMMENT{Retrieve items for all QA utterances in Sub-task A}
\STATE $\text{Ranked\_Items\_B} \gets A.\text{batch\_rank}(\text{retrieved\_items},$ \\
\hspace*{2.5em} $\text{Relevant\_Labels\_A}, \text{QA\_utterances\_B})$

\hfill \COMMENT{Rank retrieved items based on Sub-task A labels and QA utterances}
\STATE $\text{Ranked\_Items\_B} \gets \text{aggregate\_and\_finalize}(\text{Ranked\_Items\_B})$ \hfill \COMMENT{Aggregate and finalize the rankings}
\RETURN $\text{Ranked\_Items\_B}$
\end{algorithmic}
\end{algorithm}

\begin{figure*}[t]
\centering

\begin{subfigure}[t]{0.45\textwidth}
\centering
\begin{p1chat}
\scriptsize
\faUser\ [P1]: Hey there! I'm excited to explore some eco-friendly cleaning options for my kitchen. Any recommendations for effective, well-rated products that smell great?
\end{p1chat}

\begin{agentchat}
\scriptsize
\textbf{Of course! Based on your interest in gentle, skin-safe, and natural products, here are a few highly rated, eco-conscious kitchen cleaning items that smell great and get the job done:}

\vspace{2pt}
\textbf{1. Dawn Ultra Platinum Foam Dishwashing Foam (Fresh Rapids Scent)} \\
\textit{Price: \$3.69} — Absorbs 5x more grease and delivers a refreshing scent. Efficient for small-batch dishwashing with minimal water.

\vspace{2pt}
\textbf{2. OxiClean Odor Blasters Versatile Odor and Stain Remover Powder (5 lb)} \\
\textit{Price: \$11.78} — Citrus-scented and chlorine-free. Great for laundry, surfaces, and kitchen messes with eco-safe performance.

\vspace{2pt}
\textbf{3. Pure Homemaker Wool Dryer Balls (6-Pack)} \\
\textit{Price: N/A} — A reusable alternative to chemical dryer sheets. Add essential oils for natural scent while softening clothes and cutting drying time.
\end{agentchat}
\vspace{-0.5em}
\caption{Persona A (Female)}
\label{fig:personaA_taskB}
\end{subfigure}
\hspace {-0.01em}
\begin{subfigure}[t]{0.46\textwidth}
\centering
\begin{p2chat}
\scriptsize
\faUser\ [P2]: I’m looking for something to help with my joints—preferably a gentle supplement I can take daily.
\end{p2chat}

\begin{agentchat}
\scriptsize
\textbf{Absolutely! Since you’ve shown interest in health supplements and ease-of-use, here are some top-rated daily joint support products tailored for comfort and effectiveness:}

\vspace{2pt}
\textbf{1. Arazo Nutrition Glucosamine Chondroitin Turmeric MSM Boswellia (180 Tablets)} \\
\textit{Price: \$24.95} — Comprehensive joint formula supporting mobility and comfort, especially for age-related stiffness.

\vspace{2pt}
\textbf{2. Vital Vitamins Brain Booster} \\
\textit{Price: \$21.95} — While primarily for cognition, its anti-inflammatory ingredients (DMAE, Ginkgo) may support joint-related stress relief and overall wellness.

\vspace{2pt}
\textbf{3. Doctor's Best Brain Magnesium} \\
\textit{Price: \$39.29} — Shown to ease muscle tension and support neurological function—useful for cramp and joint symptom overlap.
\end{agentchat}
\vspace{-0.5em}
\caption{Persona B (Male)}
\label{fig:personaB_taskB}
\end{subfigure}
\vspace{-0.8em}
\caption{Cross-task recommendations in Session B. Each agent response reflects inferred needs from prior interactions in Task A to guide Task B recommendations that maintain coherence and personalization.}
\label{fig:recommendation_cross_task}
\end{figure*}

\noindent\textbf{Empirical Analysis}: Fig.~\ref{fig:recommendation_cross_task} presents qualitative results for two distinct personas. For Persona A, the agent succeeds in translating previously inferred values (e.g., eco-friendliness, fragrance-free preferences, budget-consciousness) into targeted recommendations for environmentally safe kitchen cleaning products. The suggestions are not only relevant but stylistically consistent with the user’s prior interaction pattern— well-rated, and naturally formulated items.

For Persona B , the performance is more varied. While the agent correctly carries over signals such as trust in established brands and interest in daily wellness routines, some recommendations show only loose alignment with the user’s new goal. For instance, productivity accessories appear alongside joint supplements—useful perhaps in general health contexts, but peripheral to the user’s expressed intent. This suggests an overextension of prior preferences without adequate goal re-grounding.

\textbf{Analytical Takeaways.} These results illuminate both the potential and limitations of LLM-driven agents in navigating intra-session personalization. When user goals remain semantically adjacent—as in the case of Persona A—carryover of contextual knowledge can enable meaningful continuity. However, for users with evolving or multidimensional goals, agents must learn to filter and reinterpret past signals rather than reapply them wholesale. More broadly, this challenge exposes the need for dynamic memory management in generative agents: the ability to store, weigh, and selectively re-engage past preferences depending on goal alignment. Simply retaining all prior information risks diluting precision. Future work should explore mechanisms—such as salience scoring, goal re-grounding prompts, or hierarchical user modeling—that help agents decide when and how to transfer knowledge across adjacent tasks.
\begin{figure}[t!]
    \centering
    \includegraphics[width=\columnwidth]{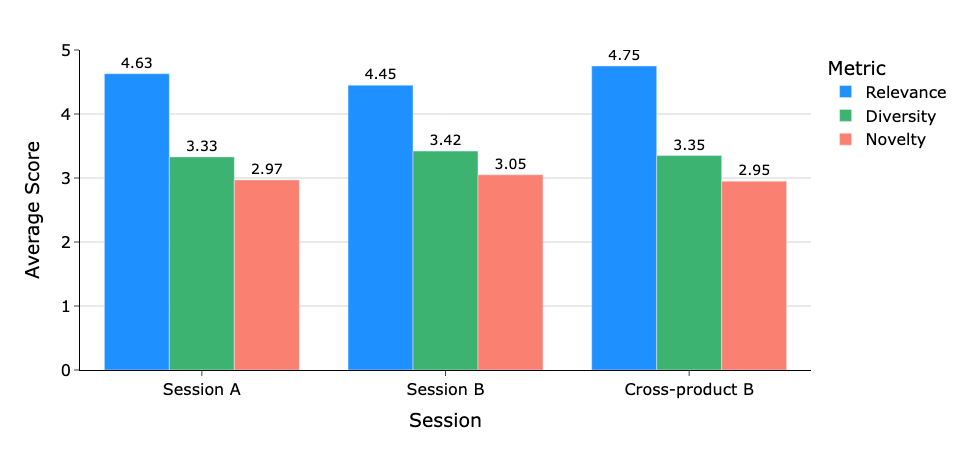}
    \vspace{-2.5em}
    \caption{Dynamic evaluation using LLM-simulated users across three personalization metrics (relevance, diversity, novelty) averaged over multiple personas and sessions. Scores range from 1 (low) to 5 (high), computed via structured LLM outputs.}
    \label{fig:dyn_evals}
    \vspace{-0.5em}
    \Description{ Illustration}
    \vspace{-1.5em}
\end{figure} 
\subsection{Step 3: Dynamic Evaluation}
Evaluating personalized agents requires a principled approach that not only examines qualitative examples but also offers quantitative metrics for broader assessment. To this end, we introduce a dynamic evaluation pipeline grounded in simulation. Specifically, we deploy a simulated user—instantiated as an LLM-driven agent (GPT-4o-mini)—to assess system outputs across multiple metrics. These evaluations are conducted \textit{across all 120 simulated users} (60 female and 60 male profiles), and the scores reported in Fig.~\ref{fig:dyn_evals} represent average values aggregated over the full user set.

While earlier sections illustrated the agent’s behavior through detailed case studies, this section aims to provide a comprehensive snapshot of personalization evaluation at scale. The evaluation is grounded in the agent’s performance across three distinct interaction contexts: (1) Session A (initial interaction), (2) Session B (follow-up session), and (3) Cross-task B (goal-shifted sub-task within the same session). These are operationalized through our structured simulation framework.Each output recommendation list is evaluated by the LLM-based simulated user using three personalization metrics:
(1.)\textbf{Relevance}: How well the items align with the user’s stated preferences. (2.) \textbf{Diversity}: Topical or functional variety among the top-$k$ items. (3.) \textbf{Novelty}: The degree to which items are unexpected or go beyond prior known preferences.

The simulated user is prompted to rate each metric on a 5-point Likert scale (1 = low, 5 = high), with justifications provided in natural language. To ensure consistency, we use deterministic prompting (temperature = 0) with fixed seeds where applicable. We also manually verified a subset of completions for coherence and fidelity. For each of the 120 user simulations, scores were computed across the three scenarios listed above, resulting in $120 \times 3 = 360$ evaluation trials. The final results presented in Fig.~\ref{fig:dyn_evals} are averaged over all trials, ensuring a stable and interpretable estimate of agent performance. Normalization was applied at the session level to mitigate variability due to differences in product space or user verbosity.

Our results reveal several key trends in agent performance across the three interaction contexts. In \textbf{Session A} (Initial Task), relevance is highest with an average score of 4.63, indicating that the agent effectively incorporates elicited preferences into its initial recommendations. However, diversity (3.33) and novelty (2.97) are comparatively lower, suggesting that the agent tends to focus narrowly on explicitly stated user needs. In \textbf{Session B}, relevance remains strong at 4.45, while diversity (3.42) and novelty (3.05) show slight improvements—reflecting more exploratory behavior as the agent gains confidence in inferred user preferences. Finally, in \textbf{Cross-task B}, relevance reaches its peak at 4.75, suggesting that the agent successfully transfers contextual knowledge from prior subtasks. Yet, novelty experiences a slight decline (2.95), indicating a stronger reliance on previously learned preferences rather than introducing new or unexpected options.

These trends reflect a desirable trade-off between exploitation and exploration. The agent is able to preserve alignment with user goals while incrementally introducing diversity and novelty over time. Nonetheless, the modest novelty scores highlight opportunities for future improvement, such as personalized diversification strategies or reinforcement-driven sampling techniques.
This section complements our qualitative demonstrations by providing large-scale evidence of the agent's adaptive capabilities. The evaluation design—combining simulated feedback, structured metrics, and session-aware aggregation—enables a robust and reproducible measure of dynamic personalization effectiveness.

\section{Conclusion and Future Work}

This paper advances a perspective on the evaluation of personalized agents that moves beyond the confines of static benchmarks and one-off metrics. Rather than introducing a fixed framework or an empirical method, we offer a reorientation in how agent adaptability should be conceptualized and assessed. By anchoring evaluation in temporally grounded simulations, structured preference elicitation, and multi-session user modeling, our work articulates a generalizable paradigm for understanding personalization as a longitudinal, interaction-driven process.

We acknowledge that our current instantiation reflects a bounded scenario—focused on two sessions and two subtasks within a single domain. While this controlled setup provides clarity and reproducibility, future extensions may involve longer temporal trajectories, multi-agent coordination, or cross-domain personalization. Furthermore, our reliance on simulated LLM-based users and structured prompting opens avenues for richer user modeling, but also limits the realism of user variability and engagement. We do not, at this stage, explore mechanisms for memory summarization, adaptive forgetting, or scalable feedback aggregation. These are not shortcomings of the paradigm itself, but practical constraints of this initial demonstration. As these components mature, we anticipate that the evaluation paradigm outlined here will become more extensible, robust, and deployable across real-world systems.

Looking forward, this perspective paves the way for a broad set of future research trajectories. First, our work highlights the need to move beyond static personalization toward proactive agent behavior, where systems can anticipate, rather than merely react to, evolving user needs. This calls for novel mechanisms that detect early signals of intent drift, calibrate proactivity to user receptiveness, and balance initiative-taking with user autonomy. Second, the role of personalization-aware memory warrants deeper investigation. Agents must not only store prior interactions, but also develop dynamic strategies for retention, abstraction, and selective forgetting—ensuring that long-term personalization remains relevant without becoming redundant or brittle. This raises open questions around memory architectures, temporal salience scoring, and hybrid episodic-semantic storage for conversational agents. Third, as personalization grows more continuous and embedded across domains, there is a pressing need for multi-session, multi-task, and multi-domain generalization benchmarks that account for sustained adaptation. Our work lays a foundation for such evaluation, but substantial datasets and methodological advances are needed to scale simulation fidelity, enable interleaved tasks, and model life-long user trajectories. Collectively, these directions underscore that personalization is not merely a model optimization challenge—it is a systems-level, longitudinal, and deeply human-centered design problem.
\begin{acks}
This work was supported by the University of Washington eScience Institute in partnership with Microsoft Azure. 
\end{acks}
\balance
\bibliographystyle{ACM-Reference-Format}
\balance
\bibliography{references}
\end{document}